\documentclass{article}
\usepackage{spconf,amsmath,epsfig,amsfonts}
\usepackage{psfrag}
\usepackage{amsmath}
\usepackage{amssymb}
\usepackage{amsfonts}
\usepackage{color}
\usepackage{tabularx}
\usepackage{subfigure}

\newcolumntype{C}[1]{>{\centering\arraybackslash}p{#1}}

\usepackage{subfigure}
\renewcommand{\thesubfigure}{\thefigure.\arabic{subfigure}}
\makeatletter
\renewcommand{\p@subfigure}{}
\renewcommand{\@thesubfigure}{{\bf Fig. \thesubfigure}.\ }
\makeatother

\def\ve#1{{\mathchoice{\mbox{\boldmath$\displaystyle #1$}}%
		      {\mbox{\boldmath$\textstyle #1$}}%
		      {\mbox{\boldmath$\scriptstyle #1$}}%
		      {\mbox{\boldmath$\scriptscriptstyle #1$}}}} 
\def\trans{\mathsf{T}}
\def\diag{\mathrm{diag}}
\newcommand{\argmax}{\mathop{\mathrm{argmax}}}
\def\PSNR{\mathrm{ PSNR}}
\def\punit{\, \mathrm}

\title{Orthogonality deficiency compensation for improved frequency selective image extrapolation}
\name{J\"urgen~Seiler, Katrin~Meisinger, and Andr\'e~Kaup}
\address{Chair of Multimedia Communications and Signal Processing, \\University of Erlangen-Nuremberg, Cauerstr. 7, 91058 Erlangen, Germany\\
{\{seiler, meisinger, kaup\}@LNT.de}}

\begin{document}
\topmargin=0mm
\ninept
\maketitle


\begin{abstract} \label{abstract}
This paper describes a very efficient algorithm for image signal extrapolation. It can be used for various applications in image and video communication, e.\ g.\ the concealment of data corrupted by transmission errors or prediction in video coding. The extrapolation is performed on a limited number of known samples and extends the signal beyond these samples. 
Therefore the signal from the known samples is iteratively projected onto different basis functions in order to generate a model of the signal. As the basis functions are not orthogonal with respect to the area of the known samples we propose a new extension, the orthogonality deficiency compensation, to cope with the non-orthogonality. Using this extension, very good extrapolation results for structured as well as for smooth areas are achievable. This algorithm improves $\PSNR$ up to $2 \punit{dB}$ and gives a better visual quality for concealment of block losses compared to extrapolation algorithms existent so far.
\end{abstract}




\section{Introduction} \label{sec:introduction} \vspace{-2mm}

Estimating data samples from known surrounding samples is a major signal processing task for modern communication applications, especially in image and video communication. The extension of a signal beyond a limited number of known samples is usually referred to as signal extrapolation. A very common task is concealment of transmission errors in mobile image or video communication by estimating lost data blocks using correctly received adjacent blocks. Another area for extrapolation is prediction whereas data samples are estimated based on known data, so only the prediction error has to be transmitted instead of the original signal.

The algorithm described in this paper is an enhancement to the frequency selective extrapolation algorithm proposed in \cite{MeK04b}. For error concealment, it was shown in \cite{MeK04b} that the original algorithm provides very good extrapolation results compared to others, such as the sequential error-concealment algorithm from Li and Orchard \cite{LiO02}, the DCT-based interpolation algorithm from \mbox{Alkachouh} and Bellanger \cite{AlB00}, the projections onto convex sets algorithm proposed by Sun and Kwok \cite{SuK95} or the maximally smooth image recovery algorithm by Wang et al. \cite{WZS93}. Even if the algorithm \cite{MeK04b} belongs to the best extra\-polation algorithms existing so far, in some rare cases it still produces visible artifacts. In addition to an increase of the $\PSNR$ our proposed algorithm produces less extrapolation artifacts. A comparison between the proposed algorithm, the original algorithm and the extrapolation algorithms mentioned above is presented at the end of this paper. Within the scope of this contribution, the algorithm is carried out for two-dimensional data sets only, but by making use of \cite{MMK06} it could easily be adapted to higher dimensional data sets as well.


\section{Signal extrapolation} \label{sec:extrapolation} \vspace{-2mm}

\begin{figure}
	\psfrag{m}{$m$}
	\psfrag{n}{$n$}
	\psfrag{M}{$M$}
	\psfrag{N}{$N$}
	\psfrag{L}{$\mathcal{L} = \mathcal{A} \cup \mathcal{B}$}
	\psfrag{A}{$\mathcal{A}$}
	\psfrag{B}{$\mathcal{B}$}
	\centering
	\includegraphics[width=0.22\textwidth]{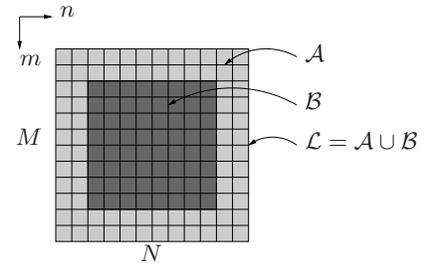}
	\caption{Data area $\mathcal{L}$ used for two-dimensional extrapolation consisting of the missing area to be estimated $\mathcal{B}$ and the known surrounding support area $\mathcal{A}$}
	\label{fig:region2d}
\end{figure}

In Fig.\ \ref{fig:region2d} a possible two-dimensional data set is shown, depicted by the two spatial dimensions $m$ and $n$. Area $\mathcal{B}$, called missing area, contains the data samples of unknown magnitude. The magnitude of these samples should be estimated by means of the data samples with known magnitude, subsumed in area $\mathcal{A}$ which is called support area. Areas $\mathcal{A}$ and $\mathcal{B}$ together form area $\mathcal{L}$ that contains all data samples involved in the extrapolation process.

\subsection{Extrapolation principle} \label{ssec:principle} 

The magnitudes of the samples in area $\mathcal{L}$ are described by the discrete function $f\left[m,n\right]$ that is defined over the whole area $\mathcal{L}$, but the magnitudes of $f\left[m,n\right]$ are only accessible over the support area $\mathcal{A}$. The algorithm aims to generate a parametric model $g\left[m,n\right]$, defined over $\mathcal{L}$ in such a way that $g\left[m,n\right]$ becomes an approximation of $f\left[m,n\right]$ over $\mathcal{A}$. The parametric model emanates from a weighted linear combination of two-dimensional basis functions $\varphi_{k} \left[m,n\right]$. The basis functions have to be mutually orthogonal with respect to the whole considered area $\mathcal{L}$. 
\begin{equation}
g\left[m,n\right] = \sum_{\forall k \in \mathcal{K}} c_k \varphi_k \left[m,n\right] ,
\end{equation}
where the weighting factors $c_k$ are denoted as expansion coefficients and the set $\mathcal{K}$ covers the indices of all basis functions used for the parametric model. In general, for decomposition of a discrete signal as many orthogonal basis functions are needed as there are samples. Therefore, the number of different orthogonal basis functions used for the generation of $g\left[m,n\right]$ equals $\left| \mathcal{L}\right|$, the number of samples in the signal. Since $\left| \mathcal{A}\right| < \left| \mathcal{L}\right|$ holds, the expansion coefficients for every basis function cannot be calculated directly and an iterative approach is needed. There, in every iteration step one basis function and the corresponding expansion coefficient is determined and added to the parametric model
\begin{equation}
g^{\left(\nu \right)} \left[m,n\right] = g^{\left(\nu-1 \right)} \left[m,n\right] + \hat{c}_u^{\left(\nu \right)} \cdot \varphi_u \left[m,n\right].
\end{equation}
Here $g^{\left(\nu \right)} \left[m,n\right]$ denotes the parametric model in the $\nu$-th iteration step, $u$ denotes the index of the basis function that was chosen for this iteration step and $\hat{c}_u^{\left(\nu \right)}$ is the estimate for the real expansion coefficient. The initial model $g^{\left(0 \right)} \left[m,n\right]$ is set to zero. In the course of the iteration a multiple selection of the same basis function is possible.

The residual approximation error $r^{\left(\nu \right)} \left[m,n\right]$ between $f\left[m,n\right]$ and the parametric model $g^{\left(\nu \right)} \left[m,n\right]$ in the $\nu$-th iteration step is expressed by
\begin{eqnarray}
r^{\left(\nu \right)} \left[m,n\right] \hspace{-0.2cm} & = & \hspace{-0.2cm} \left( f\left[m,n\right] - g^{\left(\nu \right)} \left[m,n\right] \right) \cdot b\left[m,n\right] \\
\hspace{-0.2cm} & = & \hspace{-0.2cm} \left(r^{\left(\nu - 1  \right)} \left[m,n\right] - \hat{c}_u^{\left(\nu \right)} \cdot \varphi_u \left[m,n\right]] \right) \cdot b\left[m,n\right] \! .
\end{eqnarray}
The masking function $b\left[m,n\right]$ is used to mask the missing area $\mathcal{B}$
\begin{equation}
b \left[m,n\right] = \left\{ \begin{array}{ll} 1 &,\ \forall \left(m,n\right) \in \mathcal{A} \\ 0 &,\ \forall \left(m,n\right) \in \mathcal{B} \end{array} \right. ,
\end{equation}
since the approximation error can be evaluated only over $\mathcal{A}$.

\subsection{Projection on basis functions} \label{ssec:projection}

Each iteration step begins with the weighted projection of the approximation error onto each basis function. Thereby the weighting function
\begin{equation}
w \left[m,n\right] = \left\{ \begin{array}{ll} \rho\left[m,n\right] &,\ \forall \left(m,n\right) \in \mathcal{A} \\ 0 &,\ \forall \left(m,n\right) \in \mathcal{B} \end{array} \right.
\end{equation}
is used to control the influence each sample has on the extrapolation process depending on its position in area $\mathcal{A}$. One part of $w\left[m,n\right]$ is again used to mask $\mathcal{B}$. The actual weighting is performed by $\rho\left[m,n\right]$ which can be chosen arbitrarily. A good choice for a weighting function is given in (\ref{eq:isotropic_model}).  The weighted projection onto the $k$-th basis function in the $\nu$-th iteration step results in the projection coefficient $p_k^{\left(\nu \right)}$. 
\begin{equation}
\label{eq:proj_coeff}
p_k^{\left(\nu \right)} = \frac{\displaystyle \sum_{\left(m,n\right) \in \mathcal{L}} r^{\left(\nu-1\right)} \left[m,n\right] \cdot \varphi_k \left[m,n\right] \cdot w \left[m,n\right]}{\displaystyle \sum_{\left(m,n\right) \in \mathcal{L}} w \left[m,n\right] \cdot \varphi_k^2 \left[m,n\right] } .
\end{equation}
Thereby the numerator is formed by the weighted scalar product between the approximation error and the $k$-th basis function. The denominator causes a normalization by the weighted scalar product between the selected basis function and itself.

Up to here the algorithm is comparable to the algorithm from \cite{MeK04b}. The difference between this algorithm and our proposed algorithm is that \cite{MeK04b} directly uses the projection coefficient as an estimate for the expansion coefficient without considering that the used basis functions are not mutually orthogonal any longer if evaluated with respect to the support area $\mathcal{A}$. Although the existing algorithm already provides very good extrapolation results, the consideration of the orthogonality property yields a significant enhancement as will be shown in the following.

\begin{table*}{\hfuzz=\maxdimen \newdimen\hfuzz
\hspace{1cm}\begin{tabular}{|p{6cm}|C{1.5cm}C{0.7cm}|C{1.5cm}C{0.7cm}|C{1.5cm}C{0.7cm}|} \hline
& \multicolumn{2}{c|}{``Lena''} & \multicolumn{2}{c|}{``Peppers''} & \multicolumn{2}{c|}{``Baboon''}  \\ \hline
Maximally smooth recovery \cite{WZS93}& $24.8 \punit{dB}$ && $24.6 \punit{dB}$ && $19.6 \punit{dB}$ & \\ \hline
Spatial domain interpolation \cite{AlB00} & $22.2 \punit{dB}$ && $23.4 \punit{dB}$ && $16.8 \punit{dB}$ & \\ \hline
POCS \cite{SuK95} & $22.8 \punit{dB}$ && $22.7 \punit{dB}$ && $19.0 \punit{dB}$ & \\ \hline
Sequential error concealment \cite{LiO02} & $24.7 \punit{dB}$ && $26.9 \punit{dB}$ && $18.7 \punit{dB}$ & \\ \hline
Frequency selective extrapolation \cite{MeK04b} & $24.8 \punit{dB}$ & $(15)$ & $25.3 \punit{dB}$ & $(15)$ & $19.2 \punit{dB}$ & $(5)$ \\ \hline
Presented algorithm & $26.7 \punit{dB}$ & $(500)$ & $26.8 \punit{dB}$ & $(250)$ & $19.7 \punit{dB}$ & $(200)$ \\ \hline
\end{tabular}}
\caption{Maximum achievable $\PSNR$ for different extrapolation algorithms. In brackets: number of iterations needed for the respective values for frequency selective extrapolation with and without orthogonality deficiency compensation \vspace{-0.2cm}}
\label{tab:odc_results}
\end{table*}

\subsection{Orthogonality deficiency compensation} \label{ssec:od_compensation}

Even if the basis functions are orthogonal regarding the whole area $\mathcal{L}$, the orthogonality does not apply any longer if the basis functions are evaluated with respect to the support area $\mathcal{A}$, especially if they are evaluated in combination with the weighting function. Since, depending on the relation between $\mathcal{A}$ and $\mathcal{B}$, the basis functions are still close to orthogonality this attribute should be referred to as orthogonality deficiency and not non-orthogonality.

Due to orthogonality deficiency the projection does not only lead to the contribution a basis function has to the approximation error signal but additionally contributions from other basis functions are incorporated as well. In order to determine the contribution a basis function has to the approximation signal, the expansion coefficients have to be derived from the projection coefficients. Therefore the approximation error signal is regarded as superposition of all possible basis functions, weighted with their real residual expansion coefficients
\begin{equation}
r^{\left(\nu \right)}\left[m,n\right] = \frac{1}{MN} \cdot b\left[m,n\right] \sum_{l=0,\ldots,\left|\mathcal{L}\right|-1} c_l^{\left(\nu \right)} \varphi_l \left[m,n\right] .
\end{equation}
Applying this to (\ref{eq:proj_coeff}) leads to
\[
p_k^{\left(\nu \right)} \cdot \sum_{\left(m,n\right) \in \mathcal{L}}  w \left[m,n\right] \cdot\varphi_k^2 \left[m,n\right] = \ldots \hspace{3cm}
\] \vspace{-0.5cm} \begin{equation}
\label{eq:proj_exp_coeff}
= \sum_{l=0,\ldots,\left|\mathcal{L}\right|-1} c_l^{\left(\nu \right)} \sum_{\left(m,n\right) \in \mathcal{L}} \varphi_l \left[m,n\right] \cdot \varphi_k \left[m,n\right] \cdot w \left[m,n\right] 
\end{equation}
holding for every $k=0,\ldots,\left|\mathcal{L}\right|-1$. Considering the two vectors
\begin{eqnarray}
\ve{p}^{\left(\nu\right)} &=& \left[ p_0^{\left(\nu\right)}, \ldots, p_{\left|\mathcal{L}\right|-1}^{\left(\nu\right)} \right]^\trans \\
\ve{c}^{\left(\nu\right)} &=& \left[ c_0^{\left(\nu\right)}, \ldots, c_{\left|\mathcal{L}\right|-1}^{\left(\nu\right)} \right]^\trans
\end{eqnarray}
and the matrix 
\begin{equation}
\ve{K} = \left( \begin{array}{ccc}
		\displaystyle \sum_{\left(m,n\right) \in \mathcal{L}} \tilde{w}  \tilde{\varphi}_0  \tilde{\varphi}_0  & \cdots & \displaystyle \sum_{\left(m,n\right) \in \mathcal{L}} \tilde{w}  \tilde{\varphi}_{\left|\mathcal{L}\right|-1}   \tilde{\varphi}_0  \\
		\vdots & \ddots & \vdots \\
		\displaystyle \sum_{\left(m,n\right) \in \mathcal{L}} \tilde{w}  \tilde{\varphi}_0  \tilde{\varphi}_{\left|\mathcal{L}\right|-1}  & \cdots & \displaystyle \sum_{\left(m,n\right) \in \mathcal{L}} \tilde{w}  \tilde{\varphi}_{\left|\mathcal{L}\right|-1}   \tilde{\varphi}_{\left|\mathcal{L}\right|-1} 
                \end{array} \right)
\end{equation}
with $\tilde{w} = w\left[m,n\right]$ and $\tilde{\varphi_i}= \varphi_i\left[m,n\right]$
the system of equations (\ref{eq:proj_exp_coeff}) could be rewritten in vectorial notation
\begin{equation}
\label{eq:proj_exp_coeff_vec}
\diag \left( \diag \left( \ve{K} \right) \right) \cdot \ve{p}^{\left(\nu\right)} = \ve{K} \cdot \ve{c}^{\left(\nu\right)} .
\end{equation}
The matrix $\diag \left( \diag \left( \ve{K} \right) \right)$ is a quadratic matrix with entries only on the main diagonal. The entries on the main diagonal are the same as in matrix $\ve{K}$.

Unfortunately, the matrix $\ve{K}$ cannot be inverted in general, hence an approximate solution is needed to determine the expansion coefficients from the projection coefficients. Therefore the matrix 
\begin{equation}
\hat{\ve{K}} = \left(\diag \left( \diag \left( \ve{K} \right) \right) \right)^{-1} \cdot \ve{K}
\end{equation}
is defined. For every $k$ in $0,\ldots,\left|\mathcal{L}\right|-1$ the projection coefficient $p^{\left(\nu\right)}_k$ is the sum of the real expansion coefficient $c^{\left(\nu\right)}_k$ and the terms generated by orthogonality deficiency
\begin{equation}
p^{\left(\nu\right)}_k = \hspace{-3.5mm}\sum_{l=0,\ldots,\left|\mathcal{L}\right|-1} \hspace{-4.5mm} c^{\left(\nu\right)}_l \cdot \left( \hat{\ve{K}} \right)_{k,l} =
c^{\left(\nu\right)}_k \cdot \hspace{-5mm} \sum_{l=0,\ldots,\left|\mathcal{L}\right|-1} \frac{c^{\left(\nu\right)}_l}{c^{\left(\nu\right)}_k} \cdot \left( \hat{\ve{K}} \right)_{k,l} \label{eq:proj_exp_Khat2}
\end{equation}
with $\left( \hat{\ve{K}} \right)_{k,l}$ denoting the $l$-th column in the $k$-th line of $\hat{\ve{K}}$. 

Assuming that 
\begin{equation}
\label{eq:approx_assumption}
p^{\left(\nu\right)}_k \approx \alpha c^{\left(\nu\right)}_k \ , \ \ \forall k=0,\ldots,\left|\mathcal{L}\right| -1
\end{equation}
with $\alpha$ being a constant factor, the expansion coefficient $\hat{c}^{\left(\nu\right)}_k$ can be estimated well using (\ref{eq:proj_exp_Khat2}) 
\begin{equation}
\label{eq:elaborate_approx}
\hat{c}^{\left(\nu\right)}_k = \frac{p^{\left(\nu\right)}_k}{\displaystyle \sum_{l=0,\ldots,\left|\mathcal{L}\right|-1} \frac{p^{\left(\nu\right)}_l}{p^{\left(\nu\right)}_k} \cdot \left( \hat{\ve{K}} \right)_{k,l}}
\end{equation}
Even if (\ref{eq:approx_assumption}) is a relatively weak assumption, it is valid as the summation over all projection coefficients equalizes possible minor variations of the real factor $\alpha$. 

\subsection{Basis function selection} \label{ssec:selection}

After computing the appropriate expansion coefficients for all possible basis functions, one basis function has to be chosen to be added to the parametric model in the iteration step. There are several different possible criteria for choosing the basis function. All lead to slightly different orders of the basis functions to use but also to comparable extrapolation results. The criterion used here is to choose the basis function which minimizes the distance from the approximation error signal to the weighted projection onto the basis function. The index $u$ for the basis function to use is determined by
\begin{equation}
u = \argmax_{k=0,\ldots,\left|\mathcal{L}\right|-1} \left( p_k^{\left(\nu\right)^2} \cdot \sum_{\left(m,n\right) \in \mathcal{L}} w\left[m,n\right] \cdot \varphi^2_k \left[m,n\right] \right).
\end{equation}


\section{Results}\label{sec:results} \vspace{-2mm}

\begin{figure}
	\centering
	
	\subfigure[``Lena'']{\includegraphics[width=0.3\textwidth]{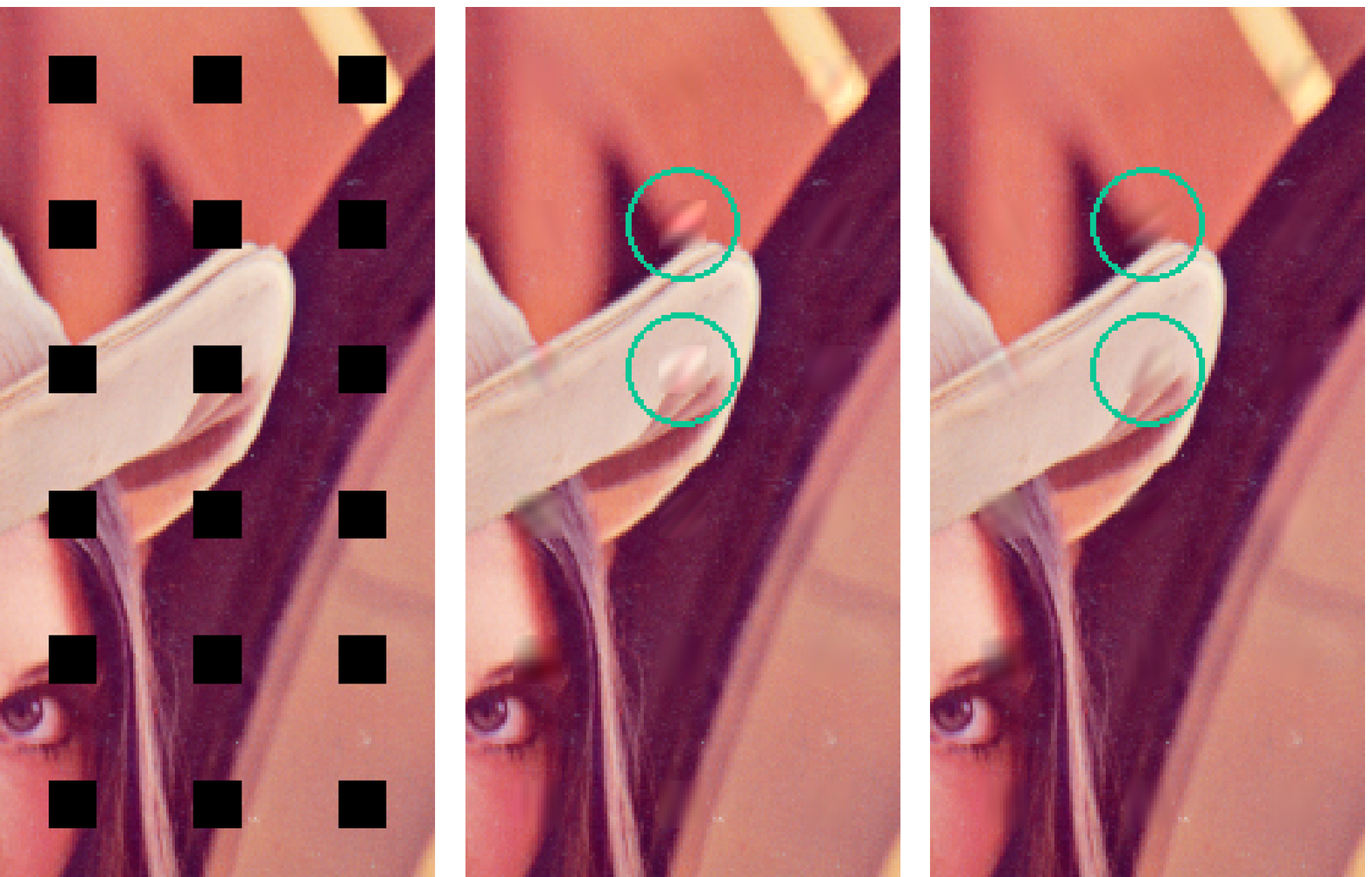} \label{fig:lena_presentation_figure}}\vspace{-0.2cm}
	\subfigure[``Peppers'']{\includegraphics[width=0.3\textwidth]{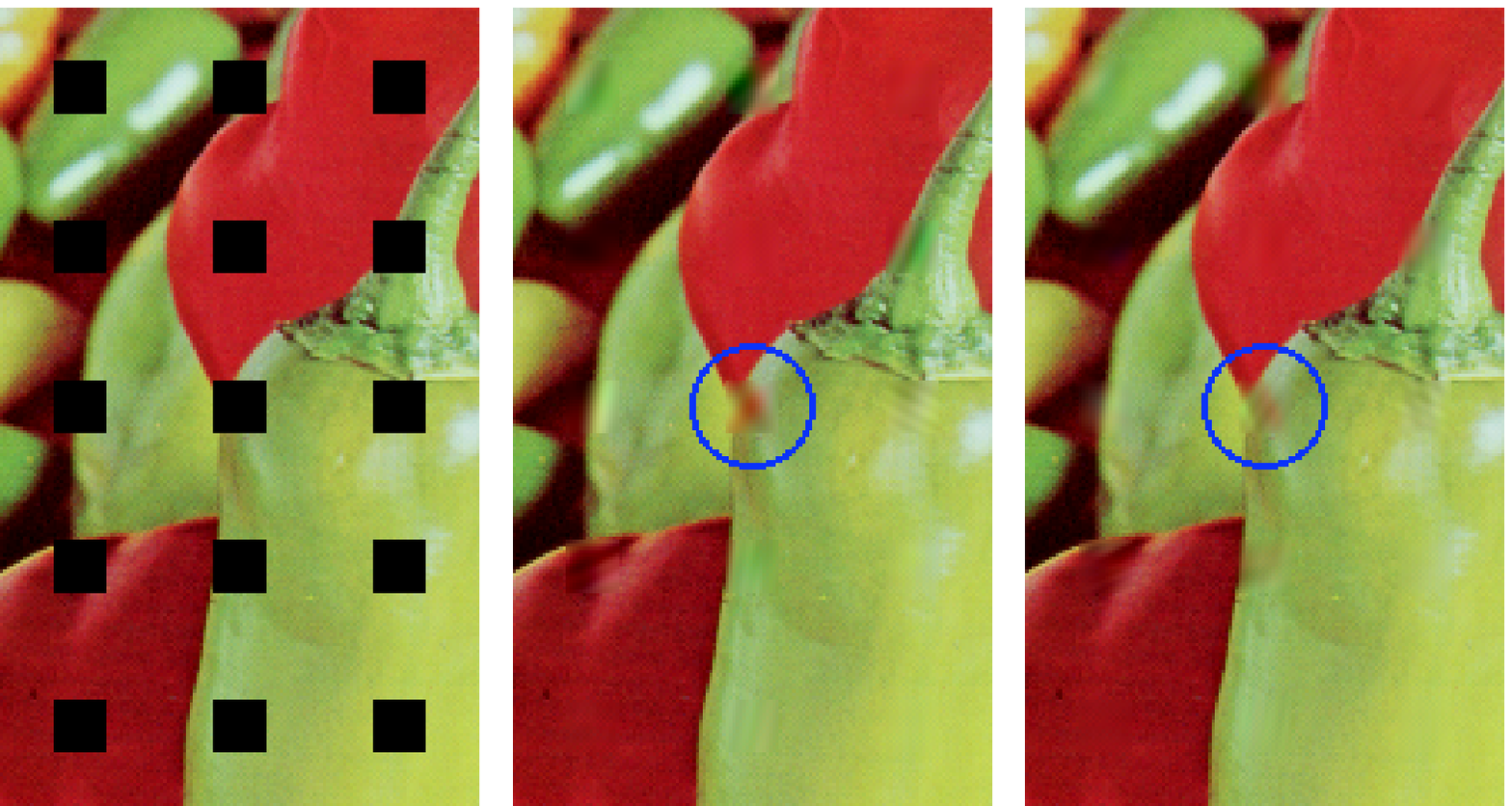} \label{fig:peppers_presentation_figure}} \vspace{-0.2cm}
	\subfigure[``Baboon'']{\includegraphics[width=0.3\textwidth]{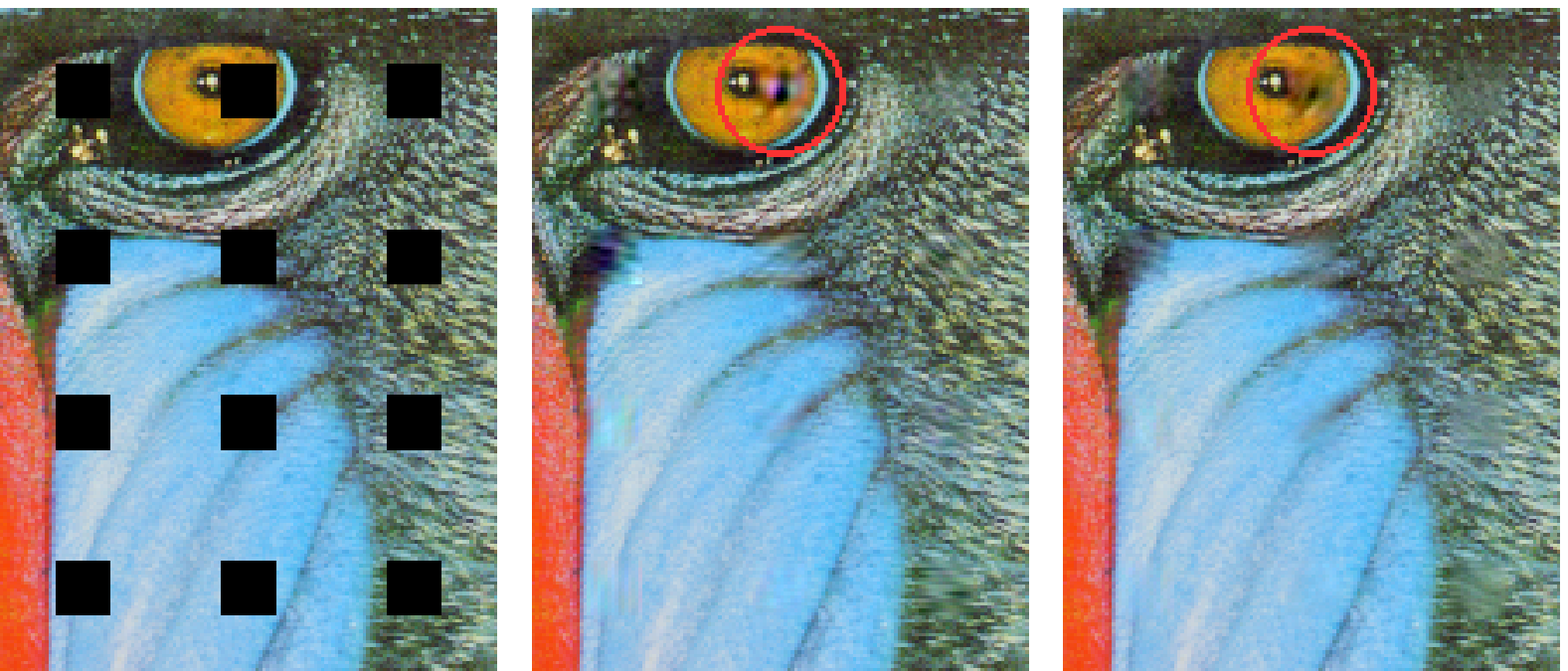} \label{fig:baboon_presentation_figure}} \vspace{-0.2cm}
	
	\caption{Error concealment with frequency selective extrapolation. Left: $16\times16$ error pattern. Mid: no orthogonality deficiency compensation used \cite{MeK04b}. Right: orthogonality deficiency compensation used. Encircled: example blocks for reduction of artifacts due to orthogonality deficiency compensation}
\end{figure}

The results for the frequency selective signal extrapolation with orthogonality deficiency compensation are shown for concealment of erroneous blocks in images. Therefore blocks are cutted out from the images ``Lena'', ``Peppers'' and ``Baboon'' according to an error pattern and the extrapolated signal is compared with the original signal in these areas. As error criterion the $\PSNR$ is evaluated for the luminance component in the lost areas. The left subfigures of Fig.\ \ref{fig:lena_presentation_figure} to Fig.\ \ref{fig:baboon_presentation_figure} illustrate some parts of the examined error pattern used for concealment of $16\times16$ pixel sized block losses.

The proposed algorithm is compared with the original frequency selective extrapolation without orthogonality deficiency compensation \cite{MeK04b} and the extrapolation algorithms from Li and Orchard \cite{LiO02}, Alkachouh and Bellanger \cite{AlB00}, Sun and Kwok \cite{SuK95} and Wang et al. \cite{WZS93}. As parameters for the frequency selective extrapolation (with and without compensation) we use ones close to the parameters proposed in \cite{MeK04b}. According to them, the weighting function used for the concealment is generated by a radial symmetric isotropic model 
\begin{equation}
\rho\left[m,n\right] = \hat{\rho}^{\sqrt{\left(m-\frac{M-1}{2}\right)^2 + \left(n-\frac{N-1}{2}\right)^2}}
\label{eq:isotropic_model}
\end{equation}
with the correlation coefficient $\hat{\rho}$ chosen to $0.8$. The support area is formed by a frame of $16$ pixels around the missing area.

\begin{figure}
\centering
\psfrag{s01}[t][t]{\color[rgb]{0,0,0}\setlength{\tabcolsep}{0pt}\begin{tabular}{c}Iterations\end{tabular}}%
\psfrag{s02}[b][b]{\color[rgb]{0,0,0}\setlength{\tabcolsep}{0pt}\begin{tabular}{c}$\PSNR$~ in $\punit{dB}$\end{tabular}}%
\psfrag{s04}[b][b]{}%
\psfrag{s06}[][]{\color[rgb]{0,0,0}\setlength{\tabcolsep}{0pt}\begin{tabular}{c} \end{tabular}}%
\psfrag{s07}[][]{\color[rgb]{0,0,0}\setlength{\tabcolsep}{0pt}\begin{tabular}{c} \end{tabular}}%
\psfrag{s17}[l][l][0.85]{\color[rgb]{0,0,0}``Lena'' without compensation}%
\psfrag{s18}[l][l][0.85]{\color[rgb]{0,0,0}``Lena'' with compensation}%
\psfrag{s19}[l][l][0.85]{\color[rgb]{0,0,0}``Peppers'' without compensation}%
\psfrag{s20}[l][l][0.85]{\color[rgb]{0,0,0}``Peppers'' with compensation}%
\psfrag{s21}[l][l][0.85]{\color[rgb]{0,0,0}``Baboon'' without compensation}%
\psfrag{s22}[l][l][0.85]{\color[rgb]{0,0,0}``Baboon'' with compensation}%
\psfrag{x12}[t][t][0.85]{$0$}%
\psfrag{x13}[t][t][0.85]{$50$}%
\psfrag{x14}[t][t][0.85]{$100$}%
\psfrag{x15}[t][t][0.85]{$150$}%
\psfrag{x16}[t][t][0.85]{$200$}%
\psfrag{x17}[t][t][0.85]{$250$}%
\psfrag{x18}[t][t][0.85]{$300$}%
\psfrag{x19}[t][t][0.85]{$350$}%
\psfrag{x20}[t][t][0.85]{$400$}%
\psfrag{x21}[t][t][0.85]{$450$}%
\psfrag{x22}[t][t][0.85]{$500$}%
\psfrag{v12}[r][r][0.85]{$12$}%
\psfrag{v13}[r][r][0.85]{$14$}%
\psfrag{v14}[r][r][0.85]{$16$}%
\psfrag{v15}[r][r][0.85]{$18$}%
\psfrag{v16}[r][r][0.85]{$20$}%
\psfrag{v17}[r][r][0.85]{$22$}%
\psfrag{v18}[r][r][0.85]{$24$}%
\psfrag{v19}[r][r][0.85]{$26$}%
\psfrag{v20}[r][r][0.85]{$28$}%
\vspace{-5mm}\includegraphics[width=0.48\textwidth]{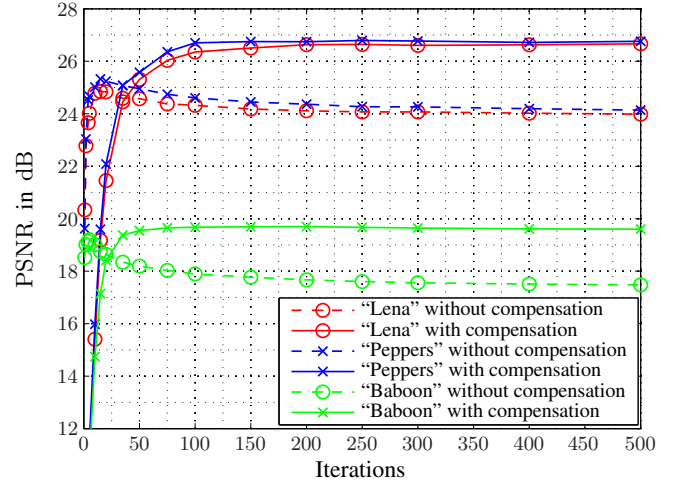}
\caption{$\PSNR$ over iterations for ``Lena'', ``Peppers'' and ``Baboon'' with losses of size $16\times16$ pixels. Solid line: orthogonality deficiency compensation used. Dashed line: original algorithm \cite{MeK04b}}
	\label{fig:results_plot}
\end{figure}

The basis functions used are the functions of the two-dimensional discrete Fourier transform whereby a very efficient realization of the algorithm presented in \cite{MeK04b} is possible. Since the entire algorithm performs in the frequency-domain only two FFTs are needed to change into the frequency domain and back respectively. Additionally, the DFT basis functions are suited especially for signal extrapolation since monotone areas, noisy regions and edges can be extrapolated very well. Here, the FFT applied has a size of $64\times64$ samples.

In Table \ref{tab:odc_results} the concealment results for the used error pattern are listed for the mentioned extrapolation algorithms. It can easily be seen that the original algorithm already outperforms most of the other methods. But in combination with the proposed orthogonality deficiency compensation a further increase of up to nearly $2 \punit{dB}$ is possible. Although the proposed algorithm performs $0.1 \punit{dB}$ worse for ``Peppers'' compared to the sequential error concealment \cite{LiO02} it provides the best extrapolation results from the regarded algorithms considering all images. 

Further the influence of the orthogonality deficiency compensation on the frequency selective extrapolation is investigated. For that in Fig.\ \ref{fig:results_plot} the $\PSNR$ achievable with and without orthogonality deficiency compensation is plotted with respect to the number of iterations. The gain in $\PSNR$ is apparent but in addition it is obvious that more iterations are needed to achieve this gain. This is caused by the circumstance that in the first iteration steps the generated parametric model gets less energy due to the orthogonality deficiency compensation. So more iterations are needed in order to gather enough energy in the model. By avoiding a too strong emphasis of single basis functions caused by orthogonality deficiency no degradation occurs with an increasing number of iterations. Whereas the original approach declines after passing a peak the proposed algorithm causes a saturation of the $\PSNR$. This is especially important as the mentioned peak normally cannot be met in practice. By using the original algorithm one always runs the risk to perform too much or not enough iterations and miss the best possible extrapolation. Due to the saturation effect this is risk is avoided using the proposed algorithm. In Table \ref{tab:odc_results} the number of iterations needed to achieve the best $\PSNR$ is listed in brackets for the two algorithms. 

One asset of the original approach is that it is able to reconstruct monotone areas as well as edges and noisy regions very well. In some rare cases artifacts are produced at sharp corners. The proposed algorithm is able to reduce these artifacts noticeably so that they are barely visible. For illustration representative parts of the examined images are shown in Fig.\ \ref{fig:lena_presentation_figure} to Fig.\ \ref{fig:baboon_presentation_figure}. On the left side, the erroneous pictures are shown. Mid, the concealed picture by use of the frequency selective extrapolation without orthogonality deficiency compensation is presented. On the right side, the ortho\-gonality deficiency compensation is used. To achieve the best visual quality the number of iterations are chosen $20$ (without compensation) and $250$ (with compensation) for ``Lena'' and ``Pepper''. For highly structured images such as ``Baboon'' more iterations are performed as indicated by $\PSNR$. This is due to the fact that structured areas have to be composed of more basis functions to give a natural appearance. So we use $50$ iterations for the uncompensated and $1000$ iterations for the compensated approach. Although the number of iterations seems to be very high far less iterations can be used as well. In practice the described algorithm only needs little more iterations than the original algorithm to outperform it (see Fig.\ \ref{fig:results_plot}). The encircled areas show some example blocks where the outlined  algorithm produces less visible artifacts compared to the original algorithm. Apparently, the proposed algorithm provides very good visual results for smooth as well as structured areas and on sharp edges.


\section{Conclusion} \label{sec:conclusion} \vspace{-2mm}

The algorithm presented here provides a very efficient approach to cope with the orthogonality deficiency that has been inherent in the frequency selective extrapolation up to now. This approach leads to very good objective and subjective results and it is an efficient method to extrapolate signals in monotone and noise-like areas and over edges. Although the proposed algorithm is computational expensive compared to the other considered algorithms, for every application a good trade-off between quality and complexity can be found by limiting the overall number of iterations for an image and distributing the iterations to blocks according to their surroundings. Nevertheless, further work will focus on deriving a simplified approach to compensate the orthogonality deficiency with less computational complexity.



\end{document}